\documentclass[aps,twocolumn,superscriptaddress,floatfix]{revtex4-1}

\usepackage{epsfig}
\usepackage{dcolumn}
\usepackage{bm}
\usepackage{amssymb}

\begin{document}

\title{The crystallization of hard disks induced by a temperature gradient}

\author{S. Wo{\l}oszczuk}
\author{A. Lipowski}
\affiliation{ Institute of Physics,
A.Mickiewicz University \\
ul.Umultowska 85,
61-614 Pozna\'n,
Poland
}

\date{\today}

\begin{abstract}
While uniform temperature has no effect on equilibrium properties of hard-core systems, its gradient might substantially change  their behaviour.  In particular,  in hard-disk system subject to temperature difference $\Delta T$ disks are repelled from the hot boundary of the system and accumulate at the cold one. Using event-driven molecular dynamics simulations we show that for sufficiently large $\Delta T$ or coverage ratio $\rho^*$, crystal forms at the cold boundary. In this spatially inhomogeneous system a significant decrease  of diffusivity of disks clearly marks the stationary interface between liquid and crystal. Such a behaviour is also supported through calculation of the radial distribution function and the bond order parameter. Simulations show that for this nonequilibrium system the equipartition of energy holds and velocity obeys the Boltzmann distribution.
\end{abstract}


\maketitle

\section{Introduction}
Hard-core systems, whose studies were initiated more than fifty years ago with a seminal papers of Alder and Wainwright \cite{ald57, ald59}, are still interesting. The  apparent simplicity of these models is very deceptive  and their behaviour even with respect to some equilibrium properties is not yet fully understood. An example of that is the nature of the liquid-crystal phase transition in the two-dimensional hard-disk system.  Although the initial proposal that this transition should be discontinuous \cite{ald62} got some support from density-functional approaches \cite{ram79,zeng90,rosenfeld90,ryzhov95}, alternative explanation challenged this possibility. In the so-called KTHNY scenario \cite{koster73,hal78,nel79,
young79,nel83,strand88} the crystal melts into a liquid via an intervening "hexatic" phase and as a result the transition consists of two continuous transitions. Recent extensive simulations support the KTHNY scenario \cite{mak06,lin06} and show Van der Waals loops, that would indicate a discontinuous transition  are only finite size effects. Although various experimental works also agree with the KTHNY scenario \cite{murray87,marcus96,kusner94,kusner95,zahn99,zahn00}, the problem is  not yet entirely settled \cite{kozak08}. 

Nonequilibrium behaviour is even more challenging especially in low dimensional systems where transport coefficients are known to diverge and the validity of the conventional hydrodynamic description is at risk \cite{pomeau75}.  Such divergences are related with the  rate of  decay of some correlation functions. The first estimations by Alder in Wainwright in 1970, suggested their slow, power-law decay \cite{ald70}, but it was initially considered with some scepticism \cite{easteal83}.  However, subsequent simulations seemed to support the slow decay and recent extensive calculations modify Alder and Wainwright estimation only by some logarithmic corrections \cite{isobe08}. 

Nonequilibrium conditions are often induced through a temperature gradient and low-dimensional hard core systems were also analysed under such conditions. In particular, the behaviour of  heat conductivity in one-dimensional systems has attracted considerable attention \cite{grass2002, deutsch2003,Lip2007} since novel, and to some extent universal, power laws  were found to describe its divergence with the system size. As another example, one can mention spatial segregation, known as  the Ludwig-Soret effect, that takes place in non mono-dispersed systems subject to temperature gradient in \cite{wieg04, lue2009}. 

Hard-core systems provide fundamental paradigms for understanding liquids, gases, glasses, crystals, colloids and even colloidal crystals. Especially the latter ones are recently of  much interest. This is because the fine control of their properties and behaviour is  now possible due to novel techniques that use  for example microgravity, dielectrophoresis, or colloidal epitaxy \cite{chaikin2003}.  Moreover, colloidal crystals obtained with such techniques  might find numerous applications as photonic crystals \cite{fan1997}, optic filters \cite{pan1997} or chemical sensors \cite{holtz1997}. 

Yet another technique used for fabrication of colloidal crystals relies on the use of the temperature gradient \cite{chaikin1999}. This intrinsically nonequilibrium technique offers fine control of the growth process and fabricated in such away crystals are relatively large (size $\sim$3mm) \cite{chaikin2000}. 
It is perhaps surprising that the temperature gradient in hard-core systems can induce crystal growth since uniform temperature is in a sense irrelevant and does not determine their phase diagram  (energy of interactions in hard-core systems is either zero or infinity and their transformations are entirely of an entropic origin). Such a counterintuitive but rather simple idea underlying the temperature-gradient method encourages numerical calculations that would support experimental findings. Such a corroboration might lead to a better understanding of the crystal growth that even in hard-core systems still offers some surprises \cite{precursor2010}. In the present paper we report  results of event-driven simulations of crystallization in a system of hard-disks subject to a temperature gradient.  In Sect. II we introduce the model and method. Results are described in Sect. III and we conclude in Sect. IV.
\section{Model and method}
In our model $N$ identical hard disks of radius $r$ and unit mass
are moving in a box of size $L_x$ and $L_y$. At the left and right ends of the channel there are thermal walls that are kept at temperatures
$T_1$ and $T_2$, respectively ($T_1>T_2$). In the thermostat we used \cite{TEHVER}, after the collision with the wall
kept at temperature $T$ a disk has its normal component sampled
from the distribution $p(v_x)=\frac{\Theta(\pm v_x)v_x}{T}{\rm
exp}(-\frac{v_x^2}{2T})$ with the sign in the argument of the
Heaviside function depending on the location of the wall. Its
parallel component is sampled from the Gaussian distribution
$p(v_y)=\frac{1}{\sqrt {2\pi T}}{\rm exp}(-\frac{v_y^2}{2T})$. In
the vertical ($y$) direction periodic boundary conditions are imposed.

An efficient way to simulate hard-disk systems is to use
event-driven molecular dynamics~\cite{RAPAPORT,BORIS,CORDERO}.
Performance of the algorithm considerably increases upon
implementing heap searching and sectorization and such methods
have already been applied to a number of
problems~\cite{miller04,Lip2006,Lip2007}.

Initially, disks are uniformly distributed inside the box. Their velocity obeys the Maxwell-Boltzmann distribution with spatially dependent temperature (i.e., kinetic energy) interpolating linearly between $T_1$ and $T_2$ (such a choice speeds up the relaxation of the systems). Usually square geometry was used ($L_x=L_y$) and the number of disks $N$ ranges from 900 (30x30) to 3600 (60x60). In some cases rectangular geometry was used with $N=2700$ (90x30) disks. Reported results do not depend significantly on the finite size effects and such modest number of disks,  in our opinion, is sufficient to analyse crystallization induced by the temperature gradient, as described in the present paper.

The behaviour of  hard-disks systems is controlled by the coverage ratio defined as
\begin{equation}
\rho^* = \frac{N\pi r^2}{L_xL_y},
\end{equation}
where $N\pi r^2$ is the total area covered by disks and $L_xL_y$ is the area of the two-dimensional simulation box.
In our simulations $\rho^*$ varied between 0.125 and 0.754. As we will show below, the temperature difference $\Delta T=T_1-T_2$ plays also an important role in determining the behaviour  in our model.
\section{Results}
Qualitatively, the evolution of our model is illustrated in Fig.{\ref{fig5}. Starting from the uniform distribution, disks are gradually expelled from the hot (left) boundary and move to the cold (right) boundary. Sufficient accumulation of disks induces crystallization (hexagonal structure). Only a fraction of disks crystallize and the hot end remains disordered. Sometimes the resulting crystal is not perfect and some vacancies form that might persist for a very long time.

\begin{figure}
\vspace{-1.1cm}
\centerline{\hspace{5cm}\epsfxsize=15cm
\epsfbox{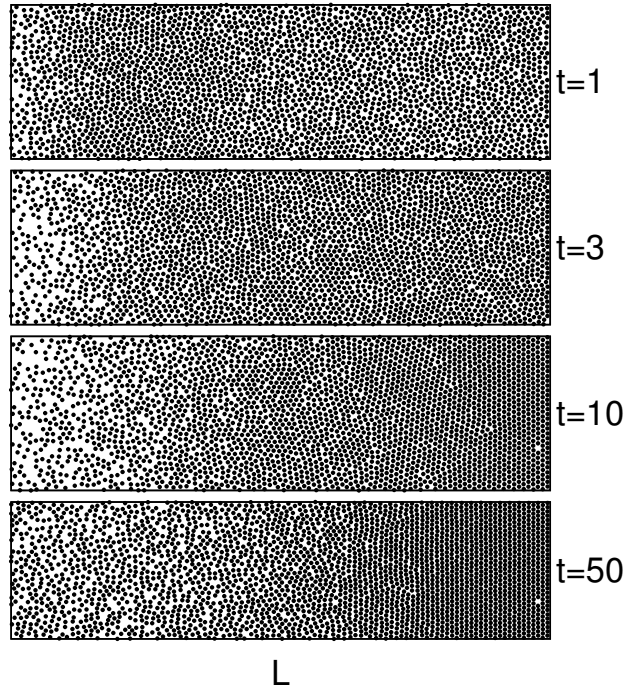} }
\caption{ \label{fig5}
Time evolution of 2700 hard disks of density $\rho^*=0.607$ simulated
in $90\times30$ box and for $T_1=5$ and $T_2=1$. Approximately at $t=10$ at the right (cold) boundary crystallization starts. Around $t=50$ the system reaches the stationary configuration. The single-site vacancy that is close to the right boundary at $t=10$ is still visible also for $t=50$.}
\end{figure}

Further evidence of crystal structure formation is seen in Fig.{\ref{timedensity}} where the spatial dependence of the average local density of disks is presented. Pronounced oscillations at the right boundary indicate formation of crystal layers.
\begin{figure}
\vspace{-0.5cm}
\centerline{\hspace{1cm}\epsfxsize=11cm
\epsfbox{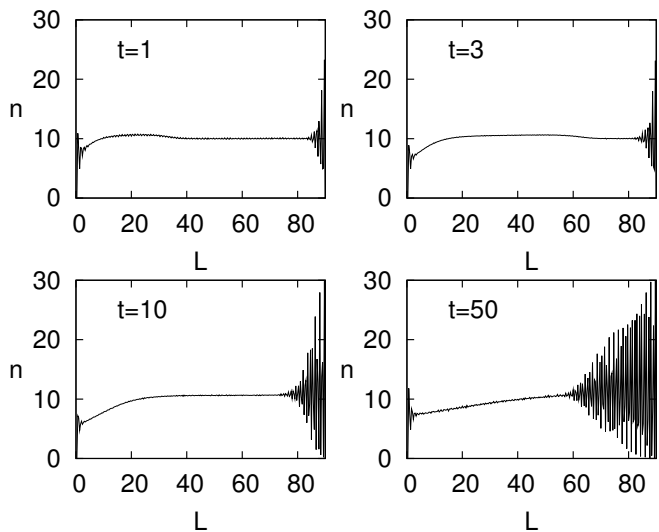} }
\caption{ \label{timedensity}
Time evolution of the spatially dependent local density. Simulations were made for the same system as in Fig.\ref{fig5} and the results are averaged over 100 independent runs.}
\end{figure}

To examine the behaviour of our model more quantitatively we measured the mean squared displacement of the disks $\langle r^2\rangle$. Since the system is spatially inhomogeneous, we divided the simulation box into several slices and calculated $\langle r^2\rangle$ and other quantities for disks in each slice separately (disks contribute to a given slice if they start in this slice). To enable comparison of diffusive properties, we fixed the simulation time as $t=10$. Such a value is much larger than mean time of ballistic (free) motion and usually smaller than the time where diffusion of disks becomes affected by finite size effects. 

In the following we discuss dependence of the diffusive behaviour on the coverage ratio $\rho^*$, the temperature difference $\Delta T$, and size effects. To further analyse  the local ordering in our system we measured the radial distribution function and the bond order parameter. Finally, we show that this strongly nonequilibrium and inhomogeneous system locally can be treated as a quasi-equilibrium one, where equipartition of energy holds and velocity of the disks obeys the Boltzmann distribution.
\subsection{Diffusion}
In Fig. \ref{fig1} we present the mean squared displacement as a function of the distance $L$ from the hot boundary. The system was divided into 30 slices and we measured separately $x$ (horizontal) and $y$ (vertical) components of the displacement. Starting from the initial configuration the system was first relaxed for a time sufficient to reach the stationary state and after that measurement of the displacement was performed. 

\begin{figure}
\vspace{9cm}
\centerline{\epsfxsize=7cm
\epsfbox{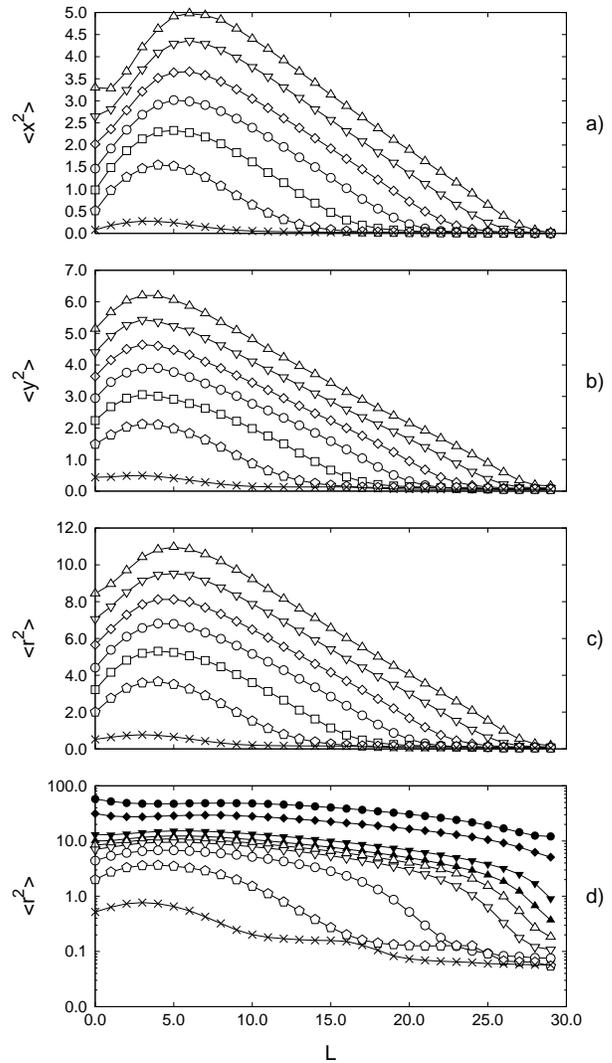} \vspace{0.5cm}}
\caption{ \label{fig1} 
Mean-squared displacement after time $t=10$ as a function of distance
from hot boundary for $N=900$,  $L_x=L_y=30$ and $T_1=5$ and $T_2=1$: (a) $x$ component $\langle x^2\rangle$
calculated for (curves from top to bottom): $\rho^*=$0.581, 0.607, 0.636, 0.664, 0.694, 0.723 and 0.754; (b) the $y$ component $\langle y^2\rangle$ for the same $\rho^*$ as in (a); (c) the isotropic displacement $\langle r^2\rangle$ for the same $\rho^*$ as in (a);  (d) the same as (c) but $\langle r^2\rangle$ is plotted on the logarithmic scale.  Top four curves are obtained for (from top) $\rho^*=0.125$, 0.282, 0.502, and 0.554. Remaining curves are obtained for the same values as in (a) but with omitted value $\rho^*=0.636$.}
\end{figure}

These data show that for sufficiently large density, diffusion of disks close to the cold boundary drops to very small values. These are the disks that form the crystal. The remaining disks diffuse freely but their diffusivity depends on their position. The closer to the hot boundary the disks are, the larger their diffusivity is. However, very close to the hot boundary this tendency reveres. Such a reduction of diffusivity is a natural consequence of a proximity to the boundary.
It is also interesting to notice that although our model is anisotropic, diffusion in the $x$ and $y$ directions is similar. 

The plot of displacement $\langle r^2\rangle$ on the logarithmic scale (Fig.\ref{fig1}, bottom panel) enabled us to approximately locate the onset of crystal formation.
Namely, $\rho^*=0.607$ seems to be a threshold value that separate data that at the cold boundary show negligible ($\sim 0.1$) displacement from those with a larger value of $\langle r^2\rangle$. We will use such a criterion for the construction of the phase diagram of the model (Fig.\ref{fig11}). Let us also notice that the threshold value $\rho^*=0.607$ is much lower than the recently reported values (0.723) for hard disks without a temperature difference \cite{mak06}.

That for hard disks and sufficiently large coverage ratio $\rho^*$ crystallization takes place is not surprising. Less obvious seems to be crystalization induced by a temperature difference. 
In Fig. \ref{fig7} we show the displacement $\langle r^2\rangle$ calculated for the fixed coverage ratio ($\rho^*=0.636$) but with varying temperature difference $\Delta T$. At the cold boundary  a transition around $T_1=3$ can be seen between low-$T_1$ (concave) and high-$T_1$ (convex) regimes that we again identify as the formation of a crystal.
\begin{figure}
\vspace{-1.6cm}
\centerline{\epsfxsize=8cm
\epsfbox{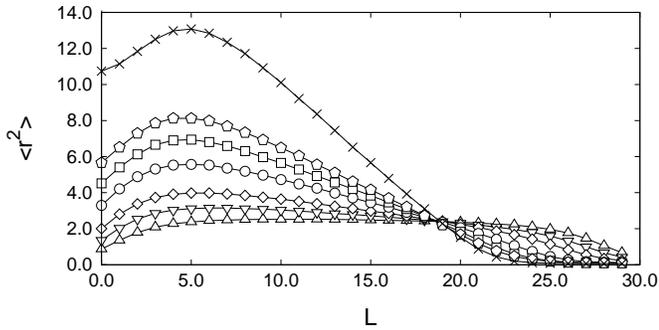} \vspace{0.5cm}}
\caption{ \label{fig7} 
The mean-square displacement $\langle r^2\rangle$  after time $t=10$ as a function of distance from the hot boundary $L$ calculated for $\rho^*=0.636$,  $N=900$, $T_2=1$, and 
$T_1$ equal to 1.1 (upward triangles), 1.5 (downward triangles), 2 (diamonds),
3 (circles), 4 (squares), 5 (pentagons) and 10 (crosses).}
\end{figure}

Let us also notice that data in Fig.\ref{fig7} for different $T_1$ intersect almost at the same point ($L=19$).  We do not understand such a behaviour yet, but assuming that for increasing $T_1$ the displacement  $\langle r^2\rangle$ will tend to the step-like function with a discontinuity at $L=19$, we can interpret this point as a high-$T_1$ limit of the existence of the crystal phase. What is in our opinion interesting, is the fact that location of this point can be estimated already from low-$T_1$ data. It is also possible that the existence of such a point  is related to the already mentioned KTHNY  scenario. Namely, moving from the cold to the hot boundary we effectively move from a crystal to a liquid phase. Our data e.g., for $T_1=10$ show that diffusion vanishes for $L=24$ and that marks the limit of existence of the crystal phase. Provided that the liquid phase terminates at $L=19$ we can speculate that for $19<L<24$ the system remains in the hexatic-like phase. Certainly, to establish if this is true or not further analysis would be needed.

Since the number of disks $N$ in simulated systems is rather modest it is important to examine the role of finite-size effects. In Fig.\ref{fig3} we present the displacement $\langle r^2\rangle$ calculated for several values of  $N$. For fixed $\rho^*$  the change of $N$ implies a change of the size $L_x(=L_y)$ and to enable comparison the horizontal axis in Fig.\ref{fig3} is rescaled. Only in the vicinity of the hot boundary there are some changes with the system size and in the rest of the system the simulation data already with $N=900$ disks seems to be sufficient. Let us notice that for fixed temperature difference ($T_1-T_2$) and increasing system size, the temperature gradient decreases. One might expect that for decreasing temperature gradient the fraction of the system that crystallizes will also decrease. Simulations show that, at least for the examined systems, this is not the case. Our results (Fig.\ref{fig3}) show fast convergence (with the system size $N$) and the resulting crystal occupies a finite fraction of the system. On the other hand, simulations with fixed temperature gradient (not presented) show that for increasing system size the size of the crystal also increases. Further analysis of the size dependence in hard disks systems with a temperature gradient would be desirable.

\begin{figure}
\vspace{2cm}
\centerline{\epsfxsize=7cm
\epsfbox{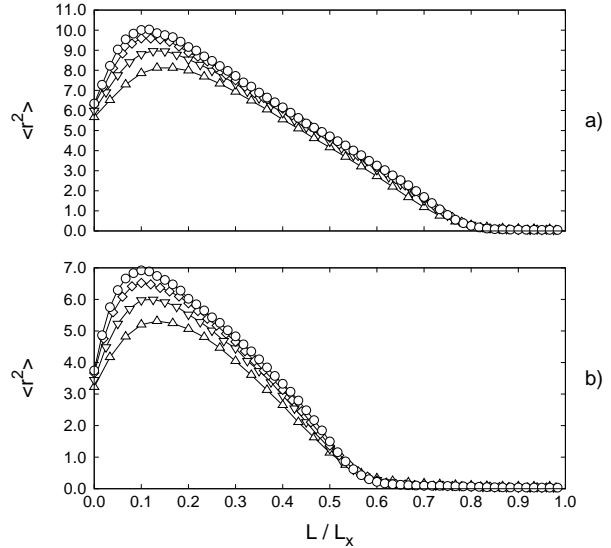} \vspace{0.5cm}}
\caption{ \label{fig3}
The mean-square displacement $\langle r^2\rangle$  after time $t=10$ as a function of the reduced distance from the hot boundary $L/L_x$ calculated for $\rho^*=0.636$ (a) and $\rho^*=0.694$ (b). Simulations were made for  $T_1=5$, $T_2=1$ and the number of disks $N=900$ (upward triangles), $N=1600$ (downward triangles), $N=2500$ (diamonds), and $N=3600$ (circles). Square geometry was used ($L_x=L_y$).
Let us notice that these data suggest the existence of a well-defined and stationary(!) boundary between crystal and liquid phases.
}
\end{figure}

\subsection{Local ordering}
Substantial change of the diffusive behaviour could indicate crystal formation. In principle however,  a drop of diffusivity might as well be due to formation of e.g., glassy phase. In the present subsection we analyse the local ordering in more details and that provide further evidence of the crystal formation.

First, we show the results of the calculation of the radial distribution function (Fig.\ref{rdf}).  At the hot part of the system ($L=0.1L_x$) only small oscillations are seen. They are more pronounced at the center ($L=0.5L_x$) and become very strong at the cold part ($L=0.9L_x$). In the latter case the split of the second peak is seen and there are some hints that its formation might be considered as a precursor of the long-range order formation (\cite{truskett98}). 

\begin{figure}
\vspace{0cm}
\centerline{\epsfxsize=9cm
\epsfbox{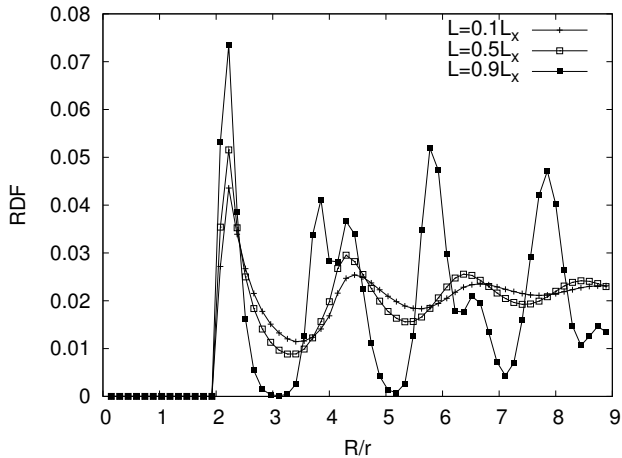} \vspace{0.0cm}}
\caption{ \label{rdf}
The radial distribution function as a function of the scaled distance $R/r$ calculated for three slabs centered around $L=0.1L_x$ (hot part), $L=0.5L_x$ (center), and $L=0.9L_x$ (cold part). Calculations were made for $\rho^*=0.636$, $T_1=5$, $T_2=1$, and $N=3600$ disks.
}
\end{figure}

Further evidence of the crystal formation comes from the calculation of the bond order parameter \cite{stein83,wang05}, $\langle q_6 \rangle$, where
the average is taken over all disks in a given slab (for a more detailed definition of $\langle q_6 \rangle$ see e.g., \cite{lechner08}).  In Fig.\ref{fig16} we present $\langle q_6 \rangle$ as a function of the reduced distance from the hot boundary $L/L_x$ calculated for $\rho^*=0.636$ (upward and downward triangles) and $\rho^*=0.694$ (diamonds and circles). Simulations were made for  $T_1=5$, $T_2=1$ and the number of disks $N=900$ (upward triangles and diamonds) and $N=1600$ (downward triangles and circles). From these data one can see that in the hot part $\langle q_6 \rangle$ is close to zero and that is the expected value in the liquid phase. In the cold part $\langle q_6 \rangle$ is much larger which is yet another indicator of the formation of crystal structure.

Let us finally point out that the region with positive diffusion (liquid) seems to be spatially well separated from the region with vanishing diffusion (crystal). Thus our model offers the possibility to study stationary properties of crystal-liquid interface, however, more detailed analysis of this issue is yet to be done.

\begin{figure}
\vspace{-1.45cm}
\centerline{\epsfxsize=7cm
\epsfbox{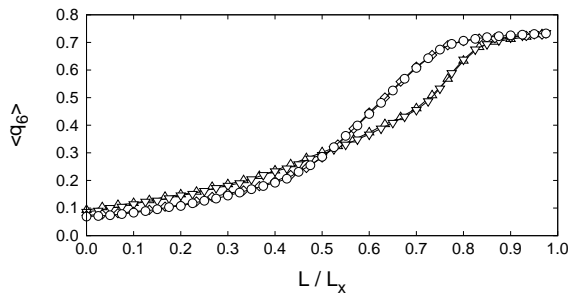} \vspace{0.5cm}}
\caption{ \label{fig16}
Mean bond order parameter, $\langle q_6 \rangle$, as a function of the reduced distance from the hot boundary $L/L_x$ calculated for $\rho^*=0.636$ (upward and downward triangles) and $\rho^*=0.694$ (diamonds and circles). Simulations were made for  $T_1=5$, $T_2=1$ and the number of disks $N=900$ (upward triangles and diamonds) and $N=1600$ (downward triangles and circles). Square geometry was used ($L_x=L_y$). The average is taken over all disks in a given sector.
}
\end{figure}

\subsection{Phase diagram}
Our main finding is that in a hard-disk system sufficiently strong temperature difference $\Delta T$ induces crystal formation. The threshold  value of $\Delta T$ depends on $\rho^*$ and the corresponding phase diagram is shown in Fig.\ref{fig11}. As a criterion for the appearance of the crystal we used the behaviour of the displacement $\langle r^2\rangle$. We considered that crystal was formed if disks in the 10\% of the system that is closest to the cold boundary had a displacement $\langle r^2\rangle$ (after time $t=10$) below 0.1. A noticeable feature is the negligible role of finite size effects and accurate location of the liquid-crystal boundary can be made already with simulations of $N=900$ disks.

One of the open problems related with this phase diagram concerns a possible existence of the hexatic phase in the $\Delta T>0$ regime, and we already mentioned such a possibility discussing Fig.\ref{fig7}.

\begin{figure}
\vspace{0.5cm}
\centerline{\epsfxsize=8cm
\epsfbox{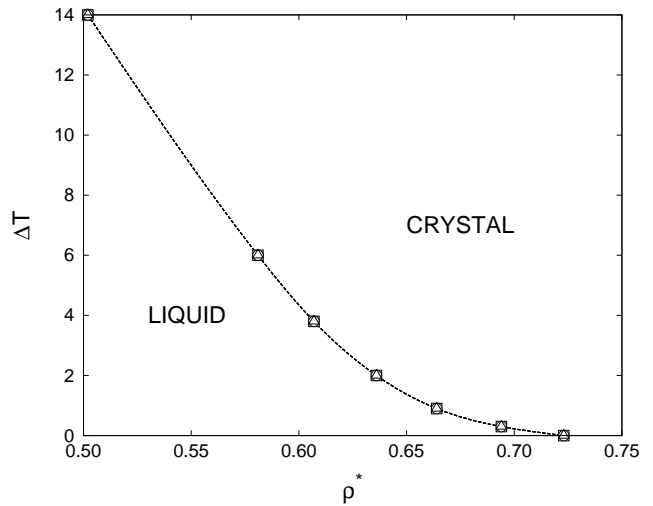} \vspace{0.7cm}}
\caption{ \label{fig11}
Phase diagram for the hard disk system of coverage ratio $\rho^*$ and with temperature difference $\Delta T$ ($T_2=1$). Simulations made for the number of disks $N$ equal to 900 (squares), 1600 (traingles), and 3600 (circles) gave nearly the same results. The onset of crystallization is detected on the value of displacement $\langle r^2\rangle$ for disks in the 10\% of the system that is closest to the cold boundary.
For a vanishing temperature difference ($\Delta T=0$) our threshold value $\rho^* \approx 0.72$, is in a good agreement with the crystallization threshold $0.723$  estimated by Kozak et al. \cite{kozak08}.
}
\end{figure}

\subsection{Quasi-equilibrium}
Although nonequilibrium systems posses a number of distinctive features, recent studies show that some notions and concepts that we use to describe equilibrium systems might be applicable also to nonequilibrium ones. Indeed, there is a number of works where nonequilibrium generalizations of entropy, temperature, or free-energy were successfully used. However, equilibrium analogies might be misleading as well. In particular, there are some indications that various, and in equilibrium equivalent, definitions of temperature in nonequilibrium yield different values \cite{crisanti}. Intensively studied systems with such ambiguities include driven granular mixtures, where each component seems to be characterized by its own temperature \cite{barrat2004}.  Moreover, in some glassy systems the effective temperature defined through fluctuation-dissipation relations turns out to depend on the choice of observables \cite{fielding2002}. Some other cornerstones of equilibrium statistical mechanics like equipartition of energy \cite {barrat2002} or gaussian distribution of velocities \cite{bennaim2002} were also reported to break down in some granular systems.

\begin{figure}
\vspace{4cm}
\centerline{\epsfxsize=7cm
\epsfbox{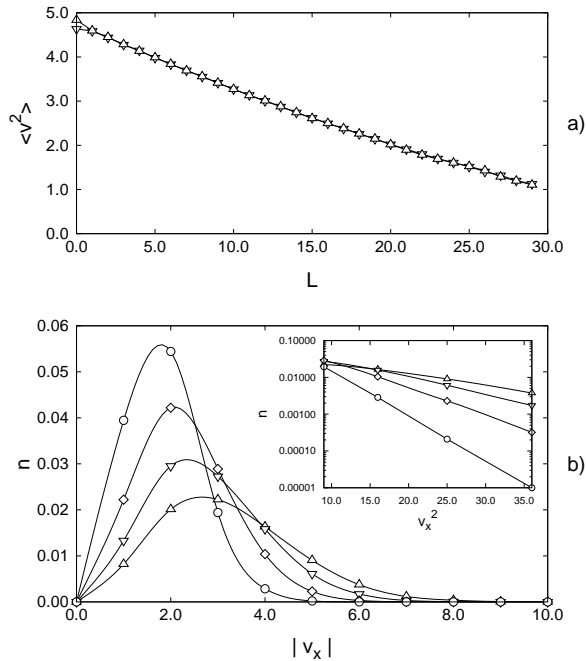} \vspace{0.5cm}}
\caption{ \label{fig6} 
(a) Average  square of horizontal (downward triangles) and vertical (upward triangles) components of velocity as a function of the distance from the hot boundary $L$. Simulations are made for $N=900$ disks, $T_1=5$  and $T_2=1$ and for coverage ratio $\rho^*=0.636$. The system is divided into 30 sectors and each point is an average for disks in a given sector. Within numerical accuracy these data support equipartition of energy in this model; (b) Distribution of $|v_x|$ in four choosen sectors located at
$L=2$ (upward triangles), $L =10)$  (downward triangles),
$L =20)$ (diamonds), and $L =28)$ (circles).
The inset shows the distribution in the semi-logarithmic plot and the linearity confirms the Boltzmann distribution.
}
\end{figure}

The nonequilibrium character of our model is caused by the temperature gradient imposed at the boundaries. Nevertheless, it is tempting to think of such a system as being in local equilibrium. In this subsection we show that this is indeed the case, at least with respect to the equipartition of energy and distribution of velocities. 
In the upper part of Fig.\ref{fig6} we present the average square of the horizontal ($\langle v_x^2\rangle$) and vertical ($\langle v_y^2\rangle$) components of the velocity. According to equipartition of energy in equilibrium systems they should be equal 
\begin{equation}
\frac{m\langle v_x^2\rangle}{2}=\frac{m\langle v_y^2\rangle}{2}=\frac{T}{2}, {\rm and} \ m=1
\label{equi}
\end{equation}
 Since our model is nonequilibrium and spatially anisotropic (heat transfer takes place only in the horizontal direction) the equality $\langle v_x^2\rangle=\langle v_y^2\rangle$ is by no means obvious (let us recall that in the perpendicular ($y$) direction we imposed periodic boundary conditions). Our simulations, however, clearly support equipartition of energy in this case (Fig.\ref{fig6}). Let us also notice that although for such a temperature difference  and coverage ratio the crystal forms at the cold (right) boundary of the system, no singularity in the behaviour of $\langle v_x^2\rangle, \langle v_y^2\rangle$ is seen (from Fig.\ref{fig3} it follows that in this case crystal exists for ($24\leq L\leq 30$). 

In the lower part of Fig.\ref{fig6} we present distribution of $|v_x|$ in four sectors (only the distribution for $v_x$ is presented and the results for  $v_y$ are qualitatively similar). The familiar Boltzmann distribution is confirmed in the inset of Fig.\ref{fig6} where the distribution is plotted in the semi-logarithmic scale. When the sectors move from left to right (from hot to cold) both the variance of the distribution and average $|v_x|$ decrease.

\section{Conclusions}
In the present paper, using event-driven molecular dynamics simulations, we examined the hard disk system with coverage ratio $\rho^*$ and subject to temperature difference $\Delta T$. Our results show that in such a system the disks accumulate at the cold boundary and for sufficiently large $\rho^*$ or $\Delta T$ they crystallize at that side. We measured diffusion of disks along the system, density profiles, radial distribution function, and bond order parameter and found that there is a clear separation of liquid and crystal. In hard-disk models, controlled only by coverage ratio $\rho^*$, the system is either in the crystal or the liquid phase and their interface usually has only a transient nature. In our model such an interface is stationary and its further analysis might provide interesting results on the coexistence of liquid and crystal. However, further analysis of this issue we left for the future. Another interesting question left for the future concerns a possible existence of the hexatic phase in hard disks with the temperature gradient. If it exists, then again, in our model we would have a possibility to study stationary coexistence of this phase with the crystal and liquid phases. It would be also desirable to extend our work to three-dimensional hard-sphere system that would be closest to experimental realizations.

\begin{acknowledgments}
This research was supported by the Ministry of Science and Higher Education Grant No. N202 071435. We gratefully
acknowledge access to the computing facilities at Pozna\'n Supercomputing and Networking Center.
\end{acknowledgments}

\bibliography{paper07}

\begin{thebibliography}{10}%
\makeatletter
\providecommand \@ifxundefined [1]{%
 \ifx #1\undefined \expandafter \@firstoftwo
 \else \expandafter \@secondoftwo
\fi
}%
\providecommand \@ifnum [1]{%
 \ifnum #1\expandafter \@firstoftwo
 \else \expandafter \@secondoftwo
\fi
}%
\providecommand \enquote [1]{``#1''}%
\providecommand \bibnamefont  [1]{#1}%
\providecommand \bibfnamefont [1]{#1}%
\providecommand \citenamefont [1]{#1}%
\providecommand\href[0]{\@sanitize\@href}%
\providecommand\@href[1]{\endgroup\@@startlink{#1}\endgroup\@@href}%
\providecommand\@@href[1]{#1\@@endlink}%
\providecommand \@sanitize [0]{\begingroup\catcode`\&12\catcode`\#12\relax}%
\@ifxundefined \pdfoutput {\@firstoftwo}{%
 \@ifnum{\z@=\pdfoutput}{\@firstoftwo}{\@secondoftwo}%
}{%
 \providecommand\@@startlink[1]{\leavevmode\special{html:<a href="#1">}}%
 \providecommand\@@endlink[0]{\special{html:</a>}}%
}{%
 \providecommand\@@startlink[1]{%
  \leavevmode
  \pdfstartlink
   attr{/Border[0 0 1 ]/H/I/C[0 1 1]}%
   user{/Subtype/Link/A<</Type/Action/S/URI/URI(#1)>>}%
  \relax
 }%
 \providecommand\@@endlink[0]{\pdfendlink}%
}%
\providecommand \url  [0]{\begingroup\@sanitize \@url }%
\providecommand \@url [1]{\endgroup\@href {#1}{\urlprefix}}%
\providecommand \urlprefix [0]{URL }%
\providecommand \Eprint[0]{\href }%
\@ifxundefined \urlstyle {%
  \providecommand \doi [1]{doi:\discretionary{}{}{}#1}%
}{%
  \providecommand \doi [0]{doi:\discretionary{}{}{}\begingroup
  \urlstyle{rm}\Url }%
}%
\providecommand \doibase [0]{http://dx.doi.org/}%
\providecommand \Doi[1]{\href{\doibase#1}}%
\providecommand \bibAnnote [3]{%
  \BibitemShut{#1}%
  \begin{quotation}\noindent
    \textsc{Key:}\ #2\\\textsc{Annotation:}\ #3%
  \end{quotation}%
}%
\providecommand \bibAnnoteFile [2]{%
  \IfFileExists{#2}{\bibAnnote {#1} {#2} {\input{#2}}}{}%
}%
\providecommand \typeout [0]{\immediate \write \m@ne }%
\providecommand \selectlanguage [0]{\@gobble}%
\providecommand \bibinfo [0]{\@secondoftwo}%
\providecommand \bibfield [0]{\@secondoftwo}%
\providecommand \translation [1]{[#1]}%
\providecommand \BibitemOpen[0]{}%
\providecommand \bibitemStop [0]{}%
\providecommand \bibitemNoStop [0]{.\EOS\space}%
\providecommand \EOS [0]{\spacefactor3000\relax}%
\providecommand \BibitemShut [1]{\csname bibitem#1\endcsname}%
\bibitem{ald57}%
  \BibitemOpen
  \bibfield{author}{%
  \bibinfo {author} {\bibfnamefont{B.~J.}\ \bibnamefont{Alder}}\ and\ \bibinfo
  {author} {\bibfnamefont{T.~E.}\ \bibnamefont{Wainwright}},\ }%
  \bibfield{journal}{%
  \bibinfo {journal} {J. Chem. Phys.}\ }%
  \textbf{\bibinfo {volume} {27}},\ \bibinfo {pages} {1208} (\bibinfo {year}
  {1957})%
  \bibAnnoteFile{NoStop}{ald57}%
\bibitem{ald59}%
  \BibitemOpen
  \bibfield{author}{%
  \bibinfo {author} {\bibfnamefont{B.~J.}\ \bibnamefont{Alder}}\ and\ \bibinfo
  {author} {\bibfnamefont{T.~E.}\ \bibnamefont{Wainwright}},\ }%
  \bibfield{journal}{%
  \bibinfo {journal} {J. Chem. Phys.}\ }%
  \textbf{\bibinfo {volume} {31}},\ \bibinfo {pages} {459} (\bibinfo {year}
  {1959})%
  \bibAnnoteFile{NoStop}{ald59}%
\bibitem{ald62}%
  \BibitemOpen
  \bibfield{author}{%
  \bibinfo {author} {\bibfnamefont{B.~J.}\ \bibnamefont{Alder}}\ and\ \bibinfo
  {author} {\bibfnamefont{T.~E.}\ \bibnamefont{Wainwright}},\ }%
  \bibfield{journal}{%
  \bibinfo {journal} {Phys. Rev.}\ }%
  \textbf{\bibinfo {volume} {127}},\ \bibinfo {pages} {359} (\bibinfo {year}
  {1962})%
  \bibAnnoteFile{NoStop}{ald62}%
\bibitem{ram79}%
  \BibitemOpen
  \bibfield{author}{%
  \bibinfo {author} {\bibfnamefont{T.~V.}\ \bibnamefont{Ramakrishnan}},\ }%
  \bibfield{journal}{%
  \bibinfo {journal} {Phys. Rev. Lett.}\ }%
  \textbf{\bibinfo {volume} {42}},\ \bibinfo {pages} {795} (\bibinfo {year}
  {1979})%
  \bibAnnoteFile{NoStop}{ram79}%
\bibitem{zeng90}%
  \BibitemOpen
  \bibfield{author}{%
  \bibinfo {author} {\bibfnamefont{X.~C.}\ \bibnamefont{Zeng}}\ and\ \bibinfo
  {author} {\bibfnamefont{D.~W.}\ \bibnamefont{Oxtoby}},\ }%
  \bibfield{journal}{%
  \bibinfo {journal} {J. Chem. Phys.}\ }%
  \textbf{\bibinfo {volume} {93}},\ \bibinfo {pages} {2692} (\bibinfo {year}
  {1990})%
  \bibAnnoteFile{NoStop}{zeng90}%
\bibitem{rosenfeld90}%
  \BibitemOpen
  \bibfield{author}{%
  \bibinfo {author} {\bibfnamefont{Y.}~\bibnamefont{Rosenfeld}},\ }%
  \bibfield{journal}{%
  \bibinfo {journal} {Phys. Rev. A}\ }%
  \textbf{\bibinfo {volume} {42}},\ \bibinfo {pages} {5978} (\bibinfo {year}
  {1990})%
  \bibAnnoteFile{NoStop}{rosenfeld90}%
\bibitem{ryzhov95}%
  \BibitemOpen
  \bibfield{author}{%
  \bibinfo {author} {\bibfnamefont{V.~N.}\ \bibnamefont{Ryzhov}}\ and\ \bibinfo
  {author} {\bibfnamefont{A.~A.}\ \bibnamefont{Tareyeva}},\ }%
  \bibfield{journal}{%
  \bibinfo {journal} {Phys. Rev. B}\ }%
  \textbf{\bibinfo {volume} {51}},\ \bibinfo {pages} {8789} (\bibinfo {year}
  {1995})%
  \bibAnnoteFile{NoStop}{ryzhov95}%
\bibitem{koster73}%
  \BibitemOpen
  \bibfield{author}{%
  \bibinfo {author} {\bibfnamefont{J.~M.}\ \bibnamefont{Kosterlitz}}\ and\
  \bibinfo {author} {\bibfnamefont{D.~J.}\ \bibnamefont{Thouless}},\ }%
  \bibfield{journal}{%
  \bibinfo {journal} {J. Phys. C: Solid State Phys.}\ }%
  \textbf{\bibinfo {volume} {6}},\ \bibinfo {pages} {1181} (\bibinfo {year}
  {1973})%
  \bibAnnoteFile{NoStop}{koster73}%
\bibitem{hal78}%
  \BibitemOpen
  \bibfield{author}{%
  \bibinfo {author} {\bibfnamefont{B.~I.}\ \bibnamefont{Halperin}}\ and\
  \bibinfo {author} {\bibfnamefont{D.~R.}\ \bibnamefont{Nelson}},\ }%
  \bibfield{journal}{%
  \bibinfo {journal} {Phys. Rev. Lett.}\ }%
  \textbf{\bibinfo {volume} {41}},\ \bibinfo {pages} {121} (\bibinfo {year}
  {1978})%
  \bibAnnoteFile{NoStop}{hal78}%
\bibitem{nel79}%
  \BibitemOpen
  \bibfield{author}{%
  \bibinfo {author} {\bibfnamefont{D.~R.}\ \bibnamefont{Nelson}}\ and\ \bibinfo
  {author} {\bibfnamefont{B.~I.}\ \bibnamefont{Halperin}},\ }%
  \bibfield{journal}{%
  \bibinfo {journal} {Phys. Rev. B}\ }%
  \textbf{\bibinfo {volume} {19}},\ \bibinfo {pages} {2457} (\bibinfo {year}
  {1979})%
  \bibAnnoteFile{NoStop}{nel79}%
\bibitem{young79}%
  \BibitemOpen
  \bibfield{author}{%
  \bibinfo {author} {\bibfnamefont{A.~P.}\ \bibnamefont{Young}},\ }%
  \bibfield{journal}{%
  \bibinfo {journal} {Phys. Rev. B}\ }%
  \textbf{\bibinfo {volume} {19}},\ \bibinfo {pages} {1855} (\bibinfo {year}
  {1979})%
  \bibAnnoteFile{NoStop}{young79}%
\bibitem{nel83}%
  \BibitemOpen
  \bibfield{author}{%
  \bibinfo {author} {\bibfnamefont{D.~R.}\ \bibnamefont{Nelson}},\ }%
  \emph{\bibinfo {title} {Phase Transitions and Critical Phenomena}},\
  Vol.~\bibinfo {volume} {7}\ (\bibinfo {publisher} {London: Academic},\
  \bibinfo {year} {1983})%
  \bibAnnoteFile{NoStop}{nel83}%
\bibitem{strand88}%
  \BibitemOpen
  \bibfield{author}{%
  \bibinfo {author} {\bibfnamefont{K.~J.}\ \bibnamefont{Strandburg}},\ }%
  \bibfield{journal}{%
  \bibinfo {journal} {Rev. Mod. Phys.}\ }%
  \textbf{\bibinfo {volume} {60}},\ \bibinfo {pages} {161} (\bibinfo {year}
  {1988})%
  \bibAnnoteFile{NoStop}{strand88}%
\bibitem{mak06}%
  \BibitemOpen
  \bibfield{author}{%
  \bibinfo {author} {\bibfnamefont{C.~H.}\ \bibnamefont{Mak}},\ }%
  \bibfield{journal}{%
  \bibinfo {journal} {Phys. Rev. E}\ }%
  \textbf{\bibinfo {volume} {73}},\ \bibinfo {pages} {065104(R)} (\bibinfo
  {year} {2006})%
  \bibAnnoteFile{NoStop}{mak06}%
\bibitem{lin06}%
  \BibitemOpen
  \bibfield{author}{%
  \bibinfo {author} {\bibfnamefont{S.~Z.}\ \bibnamefont{Lin}}, \bibinfo
  {author} {\bibfnamefont{B.}~\bibnamefont{Zheng}},\ and\ \bibinfo {author}
  {\bibfnamefont{S.}~\bibnamefont{Trimper}},\ }%
  \bibfield{journal}{%
  \bibinfo {journal} {Phys. Rev. E}\ }%
  \textbf{\bibinfo {volume} {73}},\ \bibinfo {pages} {066106} (\bibinfo {year}
  {2006})%
  \bibAnnoteFile{NoStop}{lin06}%
\bibitem{murray87}%
  \BibitemOpen
  \bibfield{author}{%
  \bibinfo {author} {\bibfnamefont{A.~C.}\ \bibnamefont{Murray}}\ and\ \bibinfo
  {author} {\bibfnamefont{D.~H.}\ \bibnamefont{van Winkle}},\ }%
  \bibfield{journal}{%
  \bibinfo {journal} {Phys. Rev. Lett.}\ }%
  \textbf{\bibinfo {volume} {58}},\ \bibinfo {pages} {1200} (\bibinfo {year}
  {1987})%
  \bibAnnoteFile{NoStop}{murray87}%
\bibitem{marcus96}%
  \BibitemOpen
  \bibfield{author}{%
  \bibinfo {author} {\bibfnamefont{A.~H.}\ \bibnamefont{Marcus}}\ and\ \bibinfo
  {author} {\bibfnamefont{S.~A.}\ \bibnamefont{Rice}},\ }%
  \bibfield{journal}{%
  \bibinfo {journal} {Phys. Rev. Lett.}\ }%
  \textbf{\bibinfo {volume} {77}},\ \bibinfo {pages} {2577} (\bibinfo {year}
  {1996})%
  \bibAnnoteFile{NoStop}{marcus96}%
\bibitem{kusner94}%
  \BibitemOpen
  \bibfield{author}{%
  \bibinfo {author} {\bibfnamefont{R.~E.}\ \bibnamefont{Kusner}}, \bibinfo
  {author} {\bibfnamefont{J.~A.}\ \bibnamefont{Mann}}, \bibinfo {author}
  {\bibfnamefont{J.}~\bibnamefont{Kerins}},\ and\ \bibinfo {author}
  {\bibfnamefont{A.~J.}\ \bibnamefont{Dahm}},\ }%
  \bibfield{journal}{%
  \bibinfo {journal} {Phys. Rev. Lett.}\ }%
  \textbf{\bibinfo {volume} {73}},\ \bibinfo {pages} {3113} (\bibinfo {year}
  {1994})%
  \bibAnnoteFile{NoStop}{kusner94}%
\bibitem{kusner95}%
  \BibitemOpen
  \bibfield{author}{%
  \bibinfo {author} {\bibfnamefont{R.~E.}\ \bibnamefont{Kusner}}, \bibinfo
  {author} {\bibfnamefont{J.~A.}\ \bibnamefont{Mann}},\ and\ \bibinfo {author}
  {\bibfnamefont{A.~J.}\ \bibnamefont{Dahm}},\ }%
  \bibfield{journal}{%
  \bibinfo {journal} {Phys. Rev. B}\ }%
  \textbf{\bibinfo {volume} {51}},\ \bibinfo {pages} {5746} (\bibinfo {year}
  {1995})%
  \bibAnnoteFile{NoStop}{kusner95}%
\bibitem{zahn99}%
  \BibitemOpen
  \bibfield{author}{%
  \bibinfo {author} {\bibfnamefont{K.}~\bibnamefont{Zahn}}, \bibinfo {author}
  {\bibfnamefont{R.}~\bibnamefont{Lenke}},\ and\ \bibinfo {author}
  {\bibfnamefont{G.}~\bibnamefont{Maret}},\ }%
  \bibfield{journal}{%
  \bibinfo {journal} {Phys. Rev. Lett.}\ }%
  \textbf{\bibinfo {volume} {82}},\ \bibinfo {pages} {2721} (\bibinfo {year}
  {1999})%
  \bibAnnoteFile{NoStop}{zahn99}%
\bibitem{zahn00}%
  \BibitemOpen
  \bibfield{author}{%
  \bibinfo {author} {\bibfnamefont{K.}~\bibnamefont{Zahn}}\ and\ \bibinfo
  {author} {\bibfnamefont{G.}~\bibnamefont{Maret}},\ }%
  \bibfield{journal}{%
  \bibinfo {journal} {Phys. Rev. Lett.}\ }%
  \textbf{\bibinfo {volume} {85}},\ \bibinfo {pages} {3656} (\bibinfo {year}
  {2000})%
  \bibAnnoteFile{NoStop}{zahn00}%
\bibitem{kozak08}%
  \BibitemOpen
  \bibfield{author}{%
  \bibinfo {author} {\bibfnamefont{J.~J.}\ \bibnamefont{Kozak}}, \bibinfo
  {author} {\bibfnamefont{J.}~\bibnamefont{Brzeziñski}},\ and\ \bibinfo
  {author} {\bibfnamefont{S.~A.}\ \bibnamefont{Rice}},\ }%
  \bibfield{journal}{%
  \bibinfo {journal} {J. Phys. Chem. B}\ }%
  \textbf{\bibinfo {volume} {112}},\ \bibinfo {pages} {16059} (\bibinfo {year}
  {2008})%
  \bibAnnoteFile{NoStop}{kozak08}%
\bibitem{pomeau75}%
  \BibitemOpen
  \bibfield{author}{%
  \bibinfo {author} {\bibfnamefont{Y.}~\bibnamefont{Pomeau}}\ and\ \bibinfo
  {author} {\bibfnamefont{P.}~\bibnamefont{R\'esibois}},\ }%
  \bibfield{journal}{%
  \bibinfo {journal} {Phys. Rep.}\ }%
  \textbf{\bibinfo {volume} {19}},\ \bibinfo {pages} {63} (\bibinfo {year}
  {1975})%
  \bibAnnoteFile{NoStop}{pomeau75}%
\bibitem{ald70}%
  \BibitemOpen
  \bibfield{author}{%
  \bibinfo {author} {\bibfnamefont{B.~J.}\ \bibnamefont{Alder}}\ and\ \bibinfo
  {author} {\bibfnamefont{T.~E.}\ \bibnamefont{Wainwright}},\ }%
  \bibfield{journal}{%
  \bibinfo {journal} {Phys. Rev. A}\ }%
  \textbf{\bibinfo {volume} {1}},\ \bibinfo {pages} {18} (\bibinfo {year}
  {1970})%
  \bibAnnoteFile{NoStop}{ald70}%
\bibitem{easteal83}%
  \BibitemOpen
  \bibfield{author}{%
  \bibinfo {author} {\bibfnamefont{A.~J.}\ \bibnamefont{Easteal}}, \bibinfo
  {author} {\bibfnamefont{L.~A.}\ \bibnamefont{Woolf}},\ and\ \bibinfo {author}
  {\bibfnamefont{D.~L.}\ \bibnamefont{Jolly}},\ }%
  \bibfield{journal}{%
  \bibinfo {journal} {Physica A}\ }%
  \textbf{\bibinfo {volume} {121}},\ \bibinfo {pages} {286} (\bibinfo {year}
  {1983})%
  \bibAnnoteFile{NoStop}{easteal83}%
\bibitem{isobe08}%
  \BibitemOpen
  \bibfield{author}{%
  \bibinfo {author} {\bibfnamefont{M.}~\bibnamefont{Isobe}},\ }%
  \bibfield{journal}{%
  \bibinfo {journal} {Phys. Rev. E}\ }%
  \textbf{\bibinfo {volume} {77}},\ \bibinfo {pages} {021201} (\bibinfo {year}
  {2008})%
  \bibAnnoteFile{NoStop}{isobe08}%
\bibitem{grass2002}%
  \BibitemOpen
  \bibfield{author}{%
  \bibinfo {author} {\bibfnamefont{P.}~\bibnamefont{Grassberger}}, \bibinfo
  {author} {\bibfnamefont{W.}~\bibnamefont{Nadler}},\ and\ \bibinfo {author}
  {\bibfnamefont{L.}~\bibnamefont{Yang}},\ }%
  \bibfield{journal}{%
  \bibinfo {journal} {Phys. Rev. Lett.}\ }%
  \textbf{\bibinfo {volume} {89}},\ \bibinfo {pages} {180601} (\bibinfo {year}
  {2002})%
  \bibAnnoteFile{NoStop}{grass2002}%
\bibitem{deutsch2003}%
  \BibitemOpen
  \bibfield{author}{%
  \bibinfo {author} {\bibfnamefont{J.~M.}\ \bibnamefont{Deutsch}}\ and\
  \bibinfo {author} {\bibfnamefont{O.}~\bibnamefont{Narayan}},\ }%
  \bibfield{journal}{%
  \bibinfo {journal} {Phys. Rev. E}\ }%
  \textbf{\bibinfo {volume} {68}},\ \bibinfo {pages} {010201} (\bibinfo {year}
  {2003})%
  \bibAnnoteFile{NoStop}{deutsch2003}%
\bibitem{Lip2007}%
  \BibitemOpen
  \bibfield{author}{%
  \bibinfo {author} {\bibfnamefont{A.}~\bibnamefont{Lipowski}}\ and\ \bibinfo
  {author} {\bibfnamefont{D.}~\bibnamefont{Lipowska}},\ }%
  \bibfield{journal}{%
  \bibinfo {journal} {Phys. Rev. E}\ }%
  \textbf{\bibinfo {volume} {75}},\ \bibinfo {pages} {052201} (\bibinfo {year}
  {2007})%
  \bibAnnoteFile{NoStop}{Lip2007}%
\bibitem{wieg04}%
  \BibitemOpen
  \bibfield{author}{%
  \bibinfo {author} {\bibfnamefont{S.}~\bibnamefont{Wiegand}},\ }%
  \bibfield{journal}{%
  \bibinfo {journal} {J. Phys.: Condens. Matter}\ }%
  \textbf{\bibinfo {volume} {16}},\ \bibinfo {pages} {R537} (\bibinfo {year}
  {2004})%
  \bibAnnoteFile{NoStop}{wieg04}%
\bibitem{lue2009}%
  \BibitemOpen
  \bibfield{author}{%
  \bibinfo {author} {\bibfnamefont{M.~N.}\ \bibnamefont{Bannermann}}\ and\
  \bibinfo {author} {\bibfnamefont{L.}~\bibnamefont{Lue}},\ }%
  \bibfield{journal}{%
  \bibinfo {journal} {J. Chem. Phys.}\ }%
  \textbf{\bibinfo {volume} {130}},\ \bibinfo {pages} {164507} (\bibinfo {year}
  {2009})%
  \bibAnnoteFile{NoStop}{lue2009}%
\bibitem{chaikin2003}%
  \BibitemOpen
  \bibfield{author}{%
  \bibinfo {author} {\bibfnamefont{M.~S.}\ \bibnamefont{et~al.}},\ }%
  \bibfield{journal}{%
  \bibinfo {journal} {J. Phys.: Condens. Matter}\ }%
  \textbf{\bibinfo {volume} {15}},\ \bibinfo {pages} {S11} (\bibinfo {year}
  {2003})%
  \bibAnnoteFile{NoStop}{chaikin2003}%
\bibitem{fan1997}%
  \BibitemOpen
  \bibfield{author}{%
  \bibinfo {author} {\bibfnamefont{J.~D.}\ \bibnamefont{Joannopoulos}},
  \bibinfo {author} {\bibfnamefont{P.~R.}\ \bibnamefont{Villeneuve}},\ and\
  \bibinfo {author} {\bibfnamefont{S.}~\bibnamefont{Fan}},\ }%
  \bibfield{journal}{%
  \bibinfo {journal} {Nature}\ }%
  \textbf{\bibinfo {volume} {386}},\ \bibinfo {pages} {143} (\bibinfo {year}
  {1997})%
  \bibAnnoteFile{NoStop}{fan1997}%
\bibitem{pan1997}%
  \BibitemOpen
  \bibfield{author}{%
  \bibinfo {author} {\bibfnamefont{G.}~\bibnamefont{Pan}}, \bibinfo {author}
  {\bibfnamefont{R.}~\bibnamefont{Kesavamoorthy}},\ and\ \bibinfo {author}
  {\bibfnamefont{S.~A.}\ \bibnamefont{Asher}},\ }%
  \bibfield{journal}{%
  \bibinfo {journal} {Phys. Rev. Lett}\ }%
  \textbf{\bibinfo {volume} {78}},\ \bibinfo {pages} {3860} (\bibinfo {year}
  {1997})%
  \bibAnnoteFile{NoStop}{pan1997}%
\bibitem{holtz1997}%
  \BibitemOpen
  \bibfield{author}{%
  \bibinfo {author} {\bibfnamefont{J.~H.}\ \bibnamefont{Holtz}}\ and\ \bibinfo
  {author} {\bibfnamefont{S.~A.}\ \bibnamefont{Asher}},\ }%
  \bibfield{journal}{%
  \bibinfo {journal} {Nature}\ }%
  \textbf{\bibinfo {volume} {389}},\ \bibinfo {pages} {829} (\bibinfo {year}
  {1997})%
  \bibAnnoteFile{NoStop}{holtz1997}%
\bibitem{chaikin1999}%
  \BibitemOpen
  \bibfield{author}{%
  \bibinfo {author} {\bibfnamefont{Z.}~\bibnamefont{Cheng}}, \bibinfo {author}
  {\bibfnamefont{W.~B.}\ \bibnamefont{Russel}},\ and\ \bibinfo {author}
  {\bibfnamefont{P.~M.}\ \bibnamefont{Chaikin}},\ }%
  \bibfield{journal}{%
  \bibinfo {journal} {Nature}\ }%
  \textbf{\bibinfo {volume} {401}},\ \bibinfo {pages} {893} (\bibinfo {year}
  {1999})%
  \bibAnnoteFile{NoStop}{chaikin1999}%
\bibitem{chaikin2000}%
  \BibitemOpen
  \bibfield{author}{%
  \bibinfo {author} {\bibfnamefont{Z.}~\bibnamefont{Cheng}}, \bibinfo {author}
  {\bibfnamefont{J.}~\bibnamefont{Zhu}}, \bibinfo {author}
  {\bibfnamefont{W.~B.}\ \bibnamefont{Russel}},\ and\ \bibinfo {author}
  {\bibfnamefont{P.}~\bibnamefont{Chaikin}},\ }%
  \bibfield{journal}{%
  \bibinfo {journal} {Phys. Rev. Lett.}\ }%
  \textbf{\bibinfo {volume} {85}},\ \bibinfo {pages} {1460} (\bibinfo {year}
  {2000})%
  \bibAnnoteFile{NoStop}{chaikin2000}%
\bibitem{precursor2010}%
  \BibitemOpen
  \bibfield{author}{%
  \bibinfo {author} {\bibfnamefont{T.}~\bibnamefont{Schilling}}, \bibinfo
  {author} {\bibfnamefont{H.~J.}\ \bibnamefont{Sch{\"o}pe}}, \bibinfo {author}
  {\bibfnamefont{M.}~\bibnamefont{Oettel}}, \bibinfo {author}
  {\bibfnamefont{G.}~\bibnamefont{Opletal}},\ and\ \bibinfo {author}
  {\bibfnamefont{I.}~\bibnamefont{Snook}},\ }%
  \enquote{\bibinfo {title} {Precursor-mediated crystallization process in
  suspensions of hard spheres},}\ \bibinfo {note} {ArXiv: 1003.2552}%
  \bibAnnoteFile{NoStop}{precursor2010}%
\bibitem{TEHVER}%
  \BibitemOpen
  \bibfield{author}{%
  \bibinfo {author} {\bibfnamefont{R.}~\bibnamefont{Tehver}}, \bibinfo {author}
  {\bibfnamefont{F.}~\bibnamefont{Toigo}}, \bibinfo {author}
  {\bibfnamefont{J.}~\bibnamefont{Koplik}},\ and\ \bibinfo {author}
  {\bibfnamefont{J.~R.}\ \bibnamefont{Banavar}},\ }%
  \bibfield{journal}{%
  \bibinfo {journal} {Phys. Rev. E}\ }%
  \textbf{\bibinfo {volume} {57}},\ \bibinfo {pages} {R17} (\bibinfo {year}
  {1998})%
  \bibAnnoteFile{NoStop}{TEHVER}%
\bibitem{RAPAPORT}%
  \BibitemOpen
  \bibfield{author}{%
  \bibinfo {author} {\bibfnamefont{D.~C.}\ \bibnamefont{Rapaport}},\ }%
  \emph{\bibinfo {title} {The Art of Molecular Dynamics Simulations}}\
  (\bibinfo {publisher} {Cambridge University Press, Cambridge},\ \bibinfo
  {year} {1995})%
  \bibAnnoteFile{NoStop}{RAPAPORT}%
\bibitem{BORIS}%
  \BibitemOpen
  \bibfield{author}{%
  \bibinfo {author} {\bibfnamefont{B.~D.}\ \bibnamefont{Lubachevsky}},\ }%
  \bibfield{journal}{%
  \bibinfo {journal} {J. Comput. Phys.}\ }%
  \textbf{\bibinfo {volume} {94}},\ \bibinfo {pages} {255} (\bibinfo {year}
  {1991})%
  \bibAnnoteFile{NoStop}{BORIS}%
\bibitem{CORDERO}%
  \BibitemOpen
  \bibfield{author}{%
  \bibinfo {author} {\bibfnamefont{M.}~\bibnamefont{Mar{\'i}n}}\ and\ \bibinfo
  {author} {\bibfnamefont{P.}~\bibnamefont{Cordero}},\ }%
  \bibfield{journal}{%
  \bibinfo {journal} {Comput. Phys. Commun.}\ }%
  \textbf{\bibinfo {volume} {92}},\ \bibinfo {pages} {214} (\bibinfo {year}
  {1995})%
  \bibAnnoteFile{NoStop}{CORDERO}%
\bibitem{miller04}%
  \BibitemOpen
  \bibfield{author}{%
  \bibinfo {author} {\bibfnamefont{S.}~\bibnamefont{Miller}}\ and\ \bibinfo
  {author} {\bibfnamefont{S.}~\bibnamefont{Luding}},\ }%
  \bibfield{journal}{%
  \bibinfo {journal} {J. Comp. Phys.}\ }%
  \textbf{\bibinfo {volume} {193(1)}},\ \bibinfo {pages} {306} (\bibinfo {year}
  {2004})%
  \bibAnnoteFile{NoStop}{miller04}%
\bibitem{Lip2006}%
  \BibitemOpen
  \bibfield{author}{%
  \bibinfo {author} {\bibfnamefont{A.}~\bibnamefont{Lipowski}}, \bibinfo
  {author} {\bibfnamefont{D.}~\bibnamefont{Lipowska}},\ and\ \bibinfo {author}
  {\bibfnamefont{A.~L.}\ \bibnamefont{Ferreira}},\ }%
  \bibfield{journal}{%
  \bibinfo {journal} {Phys. Rev. E}\ }%
  \textbf{\bibinfo {volume} {73}},\ \bibinfo {pages} {032102} (\bibinfo {year}
  {2006})%
  \bibAnnoteFile{NoStop}{Lip2006}%
\bibitem{truskett98}%
  \BibitemOpen
  \bibfield{author}{%
  \bibinfo {author} {\bibfnamefont{T.~M.}\ \bibnamefont{Truskett}}, \bibinfo
  {author} {\bibfnamefont{S.}~\bibnamefont{Torquatto}}, \bibinfo {author}
  {\bibfnamefont{S.}~\bibnamefont{Sastry}}, \bibinfo {author}
  {\bibfnamefont{P.~G.}\ \bibnamefont{Debenedetti}},\ and\ \bibinfo {author}
  {\bibfnamefont{F.~H.}\ \bibnamefont{Stillinger}},\ }%
  \bibfield{journal}{%
  \bibinfo {journal} {Phys. Rev. E}\ }%
  \textbf{\bibinfo {volume} {58}},\ \bibinfo {pages} {3083} (\bibinfo {year}
  {1998})%
  \bibAnnoteFile{NoStop}{truskett98}%
\bibitem{stein83}%
  \BibitemOpen
  \bibfield{author}{%
  \bibinfo {author} {\bibfnamefont{P.~J.}\ \bibnamefont{Steinhardt}}, \bibinfo
  {author} {\bibfnamefont{D.~R.}\ \bibnamefont{Nelson}},\ and\ \bibinfo
  {author} {\bibfnamefont{M.}~\bibnamefont{Ronchetti}},\ }%
  \bibfield{journal}{%
  \bibinfo {journal} {Phys. Rev. B}\ }%
  \textbf{\bibinfo {volume} {28(2)}},\ \bibinfo {pages} {784} (\bibinfo {year}
  {1983})%
  \bibAnnoteFile{NoStop}{stein83}%
\bibitem{wang05}%
  \BibitemOpen
  \bibfield{author}{%
  \bibinfo {author} {\bibfnamefont{Y.}~\bibnamefont{Wang}}, \bibinfo {author}
  {\bibfnamefont{S.}~\bibnamefont{Teitel}},\ and\ \bibinfo {author}
  {\bibfnamefont{C.}~\bibnamefont{Dellago}},\ }%
  \bibfield{journal}{%
  \bibinfo {journal} {J. Chem. Phys.}\ }%
  \textbf{\bibinfo {volume} {122}},\ \bibinfo {pages} {214722} (\bibinfo {year}
  {2005})%
  \bibAnnoteFile{NoStop}{wang05}%
\bibitem{lechner08}%
  \BibitemOpen
  \bibfield{author}{%
  \bibinfo {author} {\bibfnamefont{W.}~\bibnamefont{Lechner}}\ and\ \bibinfo
  {author} {\bibfnamefont{C.}~\bibnamefont{Dellago}},\ }%
  \bibfield{journal}{%
  \bibinfo {journal} {J. Chem. Phys.}\ }%
  \textbf{\bibinfo {volume} {129}},\ \bibinfo {pages} {114707} (\bibinfo {year}
  {2008})%
  \bibAnnoteFile{NoStop}{lechner08}%
\bibitem{crisanti}%
  \BibitemOpen
  \bibfield{author}{%
  \bibinfo {author} {\bibfnamefont{A.}~\bibnamefont{Crisanti}}\ and\ \bibinfo
  {author} {\bibfnamefont{F.}~\bibnamefont{Ritort}},\ }%
  \bibfield{journal}{%
  \bibinfo {journal} {J. Phys. A}\ }%
  \textbf{\bibinfo {volume} {36}},\ \bibinfo {pages} {R181} (\bibinfo {year}
  {2003})%
  \bibAnnoteFile{NoStop}{crisanti}%
\bibitem{barrat2004}%
  \BibitemOpen
  \bibfield{author}{%
  \bibinfo {author} {\bibfnamefont{A.}~\bibnamefont{Barrat}}, \bibinfo {author}
  {\bibfnamefont{V.}~\bibnamefont{Loreto}},\ and\ \bibinfo {author}
  {\bibfnamefont{A.}~\bibnamefont{Puglisi}},\ }%
  \bibfield{journal}{%
  \bibinfo {journal} {Physica A}\ }%
  \textbf{\bibinfo {volume} {334}},\ \bibinfo {pages} {513} (\bibinfo {year}
  {2004})%
  \bibAnnoteFile{NoStop}{barrat2004}%
\bibitem{fielding2002}%
  \BibitemOpen
  \bibfield{author}{%
  \bibinfo {author} {\bibfnamefont{S.}~\bibnamefont{Fielding}}\ and\ \bibinfo
  {author} {\bibfnamefont{P.}~\bibnamefont{Sollich}},\ }%
  \bibfield{journal}{%
  \bibinfo {journal} {Phys. Rev. Lett.}\ }%
  \textbf{\bibinfo {volume} {88}},\ \bibinfo {pages} {050603} (\bibinfo {year}
  {2002})%
  \bibAnnoteFile{NoStop}{fielding2002}%
\bibitem{barrat2002}%
  \BibitemOpen
  \bibfield{author}{%
  \bibinfo {author} {\bibfnamefont{A.}~\bibnamefont{Barrat}}\ and\ \bibinfo
  {author} {\bibfnamefont{E.}~\bibnamefont{Trizac}},\ }%
  \bibfield{journal}{%
  \bibinfo {journal} {Phys. Rev. E}\ }%
  \textbf{\bibinfo {volume} {66}},\ \bibinfo {pages} {051303} (\bibinfo {year}
  {2002})%
  \bibAnnoteFile{NoStop}{barrat2002}%
\bibitem{bennaim2002}%
  \BibitemOpen
  \bibfield{author}{%
  \bibinfo {author} {\bibfnamefont{P.~L.}\ \bibnamefont{Krapivsky}}\ and\
  \bibinfo {author} {\bibfnamefont{E.}~\bibnamefont{Ben-Naim}},\ }%
  \bibfield{journal}{%
  \bibinfo {journal} {J. Phys. A}\ }%
  \textbf{\bibinfo {volume} {35}},\ \bibinfo {pages} {L147} (\bibinfo {year}
  {2002})%
  \bibAnnoteFile{NoStop}{bennaim2002}%
\end{thebibliography}%
\end{document}